\documentstyle[12pt]{article}
\begin{document}
\title{
The explanation of the deformed Schild string}
\author{Rie Kuriki Maegawa 
\footnote{Internet address: maegawa@soliton.apphy.fukui-u.ac.jp}
\\
{\it Department of Applied Physics}\\
{\it Fukui University},
{\it Fukui 910-8507, Japan}}
\maketitle

\begin{abstract}
The author comments on \cite{DSchild}.
One of the deformed actions can express 
the Neveu-Schwarz-Ramond superstring 
under three gauge conditions. 
One of these depends on a matrix induced by the string
coordinate. 

\medskip 



\end{abstract}
\begin{flushright}
\end{flushright}
%
\section{The review of the Schild string and the deformation}

We summarize the properties of the Schild string and 
the results in \cite{DSchild}. 
In next section 
we will find the relation between the deformed Schild string
in \cite{DSchild} and the Neveu-Schwarz-Ramond superstring. 

The Schild string is given by the following action
\begin{eqnarray}
S_{Schild}=\int{d^2x}\left[\frac{\Gamma}{e}\{X^{\cal A},X^{\cal B}\}^2
+\Delta{e}\right],
\label{Schild0}
\end{eqnarray}
where ${\cal A}=0,1,\cdots, D-1$ 
and $X^{\cal A}$ are coordinates for a string 
propagating in D space-time dimensions \cite{Schild}\cite{Yoneya}.
$e$, which is a scaler density, is an auxiliary field. 
$\Gamma$ and $\Delta$ are arbitrary constant.
The bracket of $\{X^{\cal A},X^{\cal B}\}$ is defined by
\begin{eqnarray}
\{X^{\cal A},X^{\cal B}\}
=\epsilon^{mn}\partial_mX^{\cal A}\partial_nX^{\cal B},
\label{brak}
\end{eqnarray}
where $\epsilon^{mn}$ is $\epsilon^{mn}=-\epsilon^{nm}$ and $\epsilon^{01}=1$.

``What are the 
impediments in constructing new Schild type superstring action
from the Neveu-Schwarz-Ramond superstring ?" 
is the honest motivation of \cite{DSchild}.
Supersymmetry can be made obvious by formulating 
the theory in the super-space.
There
we deformed the action of the Schild string.
It is based on the super-space formulation 
of the Neveu-Schwarz-Ramond superstring.
The Neveu-Schwarz-Ramond superstring 
is described in N=1 superconformal flat superspace as
\begin{eqnarray}
S_{NSR}=-\frac{1}{4\pi}\int{d^2x}{d^2\theta}E
D_{\alpha}Y^{\cal A}D^{\alpha}Y_{\cal A},
\end{eqnarray}
where $D_{\alpha}$ is the covariant derivative for the local supersymmetry and 
the index $\alpha$ denote 
the chiralities of two dimensional spinor \cite{Howe2}. 
$Y^{\cal A}$ is the superfield 
\begin{eqnarray}
Y^{\cal A}=X^{\cal A}+i\bar{\theta}\psi^{\cal A}
+\frac{i}{2}\bar{\theta}\theta{B}^{\cal A},
\end{eqnarray}
where $B^{\cal A}$ is an auxiliary field which is introduced to close
the gauge algebra of the supersymmetry on the field $\psi^{\cal A}$.
$E$ is the 
superdeterminant of the supervierbein.
The supervierbein has been found in \cite{Howe},
\begin{eqnarray}
E_m{}^a=e_m{}^a+i\bar{\theta}\gamma^a\chi_m
+\frac{1}{4}\bar{\theta}\theta{e}_m{}^aA&&\nonumber\\
E_m{}^\alpha=\frac{1}{2}\chi_m{}^\alpha
+\frac{1}{2}\theta^\mu(\gamma_5)_\mu{}^\alpha\omega_m
-\frac{1}{4}\theta^\mu(\gamma_m)_\mu{}^\alpha{A}\nonumber&&\\
-\frac{3i}{16}\bar{\theta}\theta\chi_m{}^\alpha{A}
-\frac{1}{4}\bar{\theta}\theta(\gamma_m)^{\alpha\beta}\phi_\beta,&&\\
E_\mu{}^\alpha=i\theta^\lambda(\gamma^\alpha)_{\lambda\mu},&&\nonumber\\
E_\mu{}^\alpha=\delta_{\mu}{}^\alpha
-\frac{i}{8}\bar{\theta}\theta\delta_{\mu}{}^\alpha{A},&&\nonumber
\end{eqnarray}
where $e_m{}^a$ is the vierbein 
and $\chi_m{}^\alpha$ is the Rarita-Schwinger field. 
The supergauge transformation
of the $\chi_m{}^\alpha$ contains an auxiliary field A. 
By integrating the supercoordinates, we have the action
of the Neveu-Schwarz-Ramond superstring,
\begin{eqnarray}
S_{NSR}=-\frac{1}{2\pi}\int{d^2x}e(g^{mn}\partial_m
X^{\cal A}\partial_nX_{\cal A}
+i\bar\psi^{\cal A}\gamma^m\partial_m\psi_{\cal A}
-B^{\cal A}B_{\cal A}&&\nonumber\\
-i\bar{\chi}_n\gamma^m\gamma^n\psi^{\cal A}\partial_mX_{\cal A}
+\frac{1}{8}\bar{\psi}\psi\bar{\chi}_m\gamma^n\gamma^m\chi_n),
\label{NSR}
\end{eqnarray}
where $e=\det{e_m{}^a}$.
We proposed two actions by using the supervierbein.  
One of these is expressed as
\begin{eqnarray}
S_2=\int{d^2x}{d^2\theta}\left\{
\frac{\Gamma}{E}\det{h}_{\alpha\beta}+\Delta{E}\right\},
\end{eqnarray}
where 
\begin{eqnarray}
h_{\alpha\beta}=D_\alpha{Y}^{\cal C}D_\beta{Y}_{\cal C}.
\end{eqnarray}
The determinant of $h_{\alpha\beta}$ is represented as a bracket
\begin{eqnarray}
\det{h}_{\alpha\beta}=-\frac{1}{2}\left[Y^{\cal A},Y^{\cal B}\right]^2,
\end{eqnarray}
where 
\begin{eqnarray}
\left[Y^{\cal A},Y^{\cal B}\right]
=\epsilon^{\alpha\beta}D_\alpha{Y}^{\cal A}D_\beta{Y}^{\cal B}.
\end{eqnarray}
Integrating $\theta$ we obtained
\begin{eqnarray}
S_2=\int{d^2x}\frac{\Gamma}{e}
\left[\right.B_{\cal A}B^{\cal B}\bar{\psi}_{\cal B}\psi^{\cal A}
-B^2\bar{\psi}\psi+\frac{3i}{2}
B^{\cal A}\bar{\psi}\psi\bar{\chi}_m\gamma^m\psi_{\cal A}
-2B^{\cal A}\bar{\psi}_{\cal A}
\gamma^n\psi^{\cal B}\partial_n\chi_{\cal B}\nonumber&&\\
+\left(\bar{\psi}{\psi}\right)^2
\left(\frac{3i}{8}A+\frac{1}{2}g^{mn}\bar{\chi}_n\chi^m
+\frac{1}{16e}\epsilon^{nm}\bar{\chi}_m\gamma^5\chi_n\right)\nonumber&&\\
+\bar{\psi}\psi\left(\right.i\bar{\psi}^{\cal B}\gamma^m
\partial_m\psi_{\cal B}
+g^{nm}\partial_n{X}_{\cal A}\partial_m{X}^{\cal A}
-\frac{5i}{2}g^{nm}\bar{\chi}_n\psi^{\cal A}\partial_m{X}_{\cal A}\nonumber&&\\
+\frac{i}{2e}\epsilon^{ml}\partial_m{X}^{\cal A}
\bar{\psi}^{\cal A}\gamma^5\chi_l\left.\right)
-\bar{\psi}^{\cal A}\psi^{\cal B}
g^{mn}\partial_mX_{\cal A}\partial_nX_{\cal B}\nonumber&&\\
\frac{1}{e}\{X_{\cal A},X^{\cal B}\}\bar{\psi}_{\cal A}\gamma^5\psi^{\cal B}
\left.\right]
-\int{d^2x}\frac{\Delta}{2}\left(\frac{1}{2}
\epsilon^{mn}\bar{\chi}_m\gamma^5\chi_n+ieA\right).
\label{S2}
\end{eqnarray}
This is the second deformation in \cite{DSchild}. 
It is invariable 
under the volume-preserving 
diffeomorphism in the superspace,
in which the infinitesimal parameter $\xi^M$ of the super gauge transformation satisfies
\begin{eqnarray}
\partial_M\xi^M=0.
\label{APD}
\end{eqnarray}
By expanding (\ref{APD}) 
in terms of $\theta$ we found three differential 
equations, 
\begin{eqnarray}
\partial_mf^m+\frac{i}{2}\bar{\chi}_m\gamma^m\zeta=0,
\label{mf}&&\\
i\partial_m(\gamma^m\zeta)_\alpha
+\frac{i}{2}(\gamma_5\gamma^m\zeta)_\alpha\omega_m
+\frac{1}{4}\chi_{m\alpha}\bar{\zeta}\gamma^n\gamma^m\chi_n=0,\label{zeta}&&\\
\partial_m(\bar{\zeta}\gamma^n\gamma^m\chi_n)=0,\label{supermf}
\end{eqnarray}
where $f^m$ and $\zeta^\mu$ are the parameters of the two dimensional
general coordinate
transformation and the local supersymmetry, respectively. 
The solution of
(\ref{supermf}) is 
\begin{eqnarray}
\bar{\zeta}\gamma^n\gamma^m\chi_n=C{'}^m,
\label{sol2}
\end{eqnarray}
where $C{'}^m$ is an arbitrary constant. Notice that 
in the virtue of the two dimansional gamma-matrix,
$\gamma^n\gamma_m\gamma_n=0$,
(\ref{sol2}) is
invariable under the local fermionic transformation which is defined by 
the shift of the Rarita-Shwinger field as
\begin{eqnarray}
\chi_m\rightarrow\gamma_m\eta_W,
\end{eqnarray}
where $\eta_W$, which is Majorana spinor, is the infinitesimal parameter.
We easily found  
\begin{eqnarray}
\frac{\delta{S}_2}{\delta{A}}
=\frac{3i\Gamma}{8e}(\bar{\psi}\psi)^2-\frac{\Delta{i}}{2}e=0,
\label{SA}
\end{eqnarray}
in (\ref{S2}).
In the next section we rearrange the terms in (\ref{S2})
by using (\ref{SA}).

\section{The relation between the deformed Schild string 
and the Neveu-Schwarz-Ramond superstring}

The deformed action (\ref{S2}) 
is expressed as the $e=1$ gauge fixed Neveu-Schwarz-Ramond 
superstring action and 
additional terms as follows. The result is 
\begin{eqnarray}
S_2=\pm{k}\Gamma{S}_{NSR(e=1)}+K_{\pm}+U,
\label{S22}
\end{eqnarray}
where $k$ is expressed by $\Gamma$ and $\Delta$ as 
\begin{eqnarray}
k^2\equiv\frac{4\Delta}{3\Gamma}.
\end{eqnarray}
We have introduced $k$ as
\begin{eqnarray}
(\bar{\psi}\psi)^2&=&\frac{4\Delta}{3\Gamma}e^2.
\end{eqnarray}
As the result 
of (\ref{SA})
(\ref{S22}) is explicitly written by 
\begin{eqnarray}
S_{NSR(e=1)}&=&\int{d^2x}\left[\right.g^{nm}\partial_nX^{\cal A}\partial_mX^{\cal A}
+i\bar{\psi}^{\cal B}\gamma^m\partial_m\psi_{\cal B}-B^2\nonumber\\
&&-i\bar{\chi}_l\gamma^m\gamma^l\psi^{\cal A}\partial_mX_{\cal A}
+\frac{1}{8}\bar{\chi}_m\gamma^n\gamma^m\chi_n\bar{\psi}\psi\left.\right].
\end{eqnarray}
\begin{eqnarray}
K_{\pm}=\int{d^2x}\left[\frac{\Delta}{2}e\bar{\chi}_n
\gamma^n\gamma^m\chi_m\mp\frac{3}{2}ik\Gamma
\partial_mX^{\cal A}\bar{\psi}^{\cal A}\gamma^m\gamma^n\chi_n\right]
\label{Kpm}
\end{eqnarray}
\begin{eqnarray}
U=\int{d^2x}\frac{\Gamma}{e}\{
B_{\cal A}B^{\cal B}\bar{\psi}_{\cal B}\psi^{\cal A}
\pm\frac{3i}{2}B^{\cal A}e\bar{\chi}_m\gamma^m\psi_{\cal A}
-2B^{\cal A}\bar{\psi}_{\cal A}\gamma^n\psi^{\cal B}\partial_nX_{\cal B}\nonumber&&\\
-\bar{\psi}^{\cal A}\psi^{\cal B}g^{mn}\partial_mX_{\cal A}\partial_nX_{\cal B}
+\frac{1}{e}\{X_{\cal A},X_{\cal B}\}\bar{\psi}_{\cal A}\gamma_5\psi_{\cal B}
\}.
\label{U}
\end{eqnarray}
In the calculation we have used 
\begin{eqnarray}
\gamma^m\gamma^n=g^{mn}-\frac{1}{e}\epsilon^{mn}\gamma^5,
\label{gamma}
\end{eqnarray}
where $\gamma^m=\gamma^a{e}_a{}^m$.
If we define ``a matrix like gamma-matrix'' as follows,
\begin{eqnarray}
\Gamma_{\cal A}\Gamma_{\cal B}
&\equiv&G_{{\cal A}{\cal B}}-\frac{1}{e}\Upsilon_{{\cal A}{\cal B}}\gamma^5,\nonumber\\
G_{{\cal A}{\cal B}}&\equiv&g^{mn}\partial_mX_{\cal A}\partial_nX_{\cal B}
+B_{\cal A}B_{\cal B},\label{induce metric}\\
\Upsilon_{{\cal A}{\cal B}}&\equiv&\{X_{\cal A},X_{\cal B}\},
\nonumber
\end{eqnarray}
then
we can express (\ref{U}) as
\begin{eqnarray}
U=\int{d^2x}\frac{\Gamma}{e}\{
-\bar{\psi}^{\cal A}\Gamma_{\cal A}\Gamma_{\cal B}\psi^{\cal B}
\pm\frac{3i}{2}B^{\cal A}e\bar{\chi}_m\gamma^m\psi^{\cal A}
-2B^{\cal A}\gamma^n\psi^{\cal B}\partial_n{X}_{\cal B}\}.
\end{eqnarray}
The construction in (\ref{induce metric}) is inspired by (\ref{gamma}).
We pay attention that the index $n,m$ of the gamma-matrix represents the two-dimensional
surface
of string world-sheet and the index ${\cal A},{\cal B}$ of $\Gamma_{\cal A}$ denotes the
space-time dimension where a string is propagating.
That we can neatly reexpress (\ref{S2}) by introducing a matrix defined in the space-time
is interesting point here.   

Finally we find that if we impose gauge conditions 
\begin{eqnarray}
\gamma^m\chi_m=0, 
\label{gach}
\end{eqnarray}
\begin{eqnarray}
\Gamma_{\cal A}\psi^{\cal A}=0,
\label{GAGB}
\end{eqnarray}
and 
\begin{eqnarray}
B^{\cal A}=0,
\label{ba}
\end{eqnarray}
then the deformed action (\ref{S2}) is equal to  
the $e=1$ gauge fixed Neveu-Schwarz-Ramond superstring action.
Notice that 
the third gauge condition is not imposed as the equation of motion $B^{\cal A}$
derived from (\ref{S2}).

In the last section we sum up \cite{Ka} 
in where the author treats similar gauge conditions. Then we
discuss our gauge conditions. 

\section{Discussion and summary}

In \cite{Ka} the author considers the super-Weyl invariant regularization of the 
two-dimmensional supergravity. He uses the superfield formulation
and he finally constructs the super-Liouville action
which possesses the super-Weyl invariance (the Weyl invariance and the
local fermionic symmetry) and the super-area preserving 
diffeomorphism. To define which part of the two-dimensional superdiffeomorphism group is 
compatible with the super-Weyl symmetry, he searchs
the two dimensional diffeomorphism and the local supersymmetry  
which preserve the 
gauge conditions 
\begin{eqnarray}
e=1, \hspace{0.5cm} {\rm and}\hspace{0.5cm} \gamma^m\chi_m=0.
\nonumber
\end{eqnarray}
Finally he found the constraints which the parameters of the supersummetry
and the two dimensional diffeomorphism should satisfy. 
In our case (\ref{S22})
becomes the $e=1$ gauge fixed Neveu-Schwarz-Ramond superstring action,
which is invariable under the area-preserving diffeomorphism, the local fermionic transformation
and the restricted supersymmetry,
by virtue of (\ref{gach}), (\ref{GAGB}) and (\ref{ba}).
Moreover we should notice that (\ref{mf})
represents the area-preserving diffeomorphism
\begin{eqnarray}
\partial_m{f}^m=0,
\label{mf0}
\end{eqnarray}
under $\gamma^m\chi_m=0$. 
(\ref{mf0}) is one of the constraints which he found in \cite{Ka}.
Another he found is similar to (\ref{zeta}). 

In conclusion,
the superfield formulation of superstring and supergravity
is quite interesting and usuful. In the superspace 
we can manifestly keep the supersymmmetry. 
However it is not trivial to define the 
supersymmetry which is compatible with other symmetires.   
We expect that the direction of our Schild string deformation
which preserves the global structure of the bracket (\ref{brak})
would possess the dynamics of full superstring.   
     
The author is grateful to the applied physics department of Fukui university
for its gracious hospitality.

\end{document}